\documentclass[draftcls, 12pt, onecolumn]{IEEEtran}
\IEEEoverridecommandlockouts
\usepackage{cite}
\usepackage{bm}
\usepackage{amsmath,amssymb,amsfonts}
\usepackage{algorithmic}
\usepackage{graphicx}
\usepackage{textcomp}
\usepackage{xcolor}
\usepackage{subfigure}
\def\BibTeX{{\rm B\kern-.05em{\sc i\kern-.025em b}\kern-.08em
    T\kern-.1667em\lower.7ex\hbox{E}\kern-.125emX}}

\begin{document}

\title{Spectral Efficiency Analysis  of Multi-Cell Massive MIMO Systems with Ricean Fading}

\author{
Pei Liu,
Kai Luo,
Da Chen,
Tao Jiang, 
and Michail Matthaiou

\thanks{P. Liu, K. Luo, D. Chen, and T. Jiang are with the School
of Electronic Information and Communications, Huazhong University of Science and Technology, Wuhan 430074, China (e-mail: \{peil, kluo, chenda, taojiang\}@hust.edu.cn).}
\thanks{M. Matthaiou is with the ¡ìInstitute of Electronics, Communications and Information Technology (ECIT), Queen's University Belfast, BT3 9DT, Belfast, United Kingdom (e-mail: m.matthaiou@qub.ac.uk).}
}


\maketitle

\begin{abstract}
This paper investigates the  spectral efficiency of  multi-cell massive multiple-input multiple-output systems with Ricean fading that utilize the linear maximal-ratio combining detector.
We firstly present  closed-form expressions for the effective signal-to-interference-plus-noise ratio (SINR)  with the least squares  and minimum mean squared error (MMSE) estimation methods, respectively, which apply for any number of base-station antennas $M$ and any Ricean $K$-factor. Also, the obtained results can be particularized in Rayleigh fading conditions when the Ricean $K$-factor is equal to zero.
In the following, novel exact asymptotic expressions of the effective SINR are derived in the high $M$ and high Ricean $K$-factor regimes. The corresponding analysis shows that pilot contamination is removed by the MMSE estimator when we consider both infinite $M$ and infinite Ricean $K$-factor, while the  pilot contamination phenomenon persists for the rest of cases. All the theoretical results are verified via  Monte-Carlo simulations.
\end{abstract}


\section{Introduction}
In the design of future communication systems, spectral efficiency (SE) becomes one of the dominant targets \cite{Qualcomm}. Massive multiple-input multiple-output (MIMO) \cite{Marzetta10}, where the base station (BS) is equipped with  hundreds of antennas to serve tens of users in the same time-frequency resource block, has attracted substantial attention from  academia and industry thanks to the the high SE gains provided by the massive array \cite{Ngo13,Liu17,Liu18}.  The fundamental problem of placing a massive number of antennas in a confined space, can be addressed by pushing the operating frequency in the milimeter wave band, where the wavelengths become inherently small.  We recall that mm-wave channels are typically modeled via the Ricean fading distribution due to the presence of line-of-sight (LOS) or specular components \cite{Zhang14}. Hence, the SE analysis in massive MIMO systems with Ricean fading becomes a problem of practical relevance.

We will now review the relevant state-of-the-art in the massive MIMO literature.
We firstly note the work of \cite{Ngo13}, which derived useful lower bounds on the uplink ergodic rate for different classical linear receivers and analyzed the SE performance over a single-cell massive MIMO system operating in Rayleigh fading for perfect channel state information (CSI) and imperfect CSI, respectively. The same authors extended part of the above results  into multi-cell systems and also obtained the corresponding lower bounds in \cite{Ngo132}. The authors in \cite{Bjornson16} provided a closed-form expression of the SE, explored the SE maximizing problem in multi-cell systems, and proposed the optimal system parameters. Recently, \cite{Liu17} considered the achievable uplink rates based on both the least squares (LS) and minimum mean squared error (MMSE) estimation methods in multi-cell massive MIMO systems and investigated the pilot power allocation scheme to maximize the minimum SE. As a general comment, most issues pertaining to the achievable SE of massive MIMO, massive MIMO systems in Rayleigh fading have been largely and extensively characterized. Apart from that, some recent investigations have analyzed the SE performance in massive MIMO systems based on Ricean fading. For example, for a single-cell environment, \cite{Zhang14} examined the scaling law and obtained the approximate expressions of uplink ergodic rate by considering  Ricean fading and both perfect CSI and imperfect CSI.
Moreover, for multi-cell massive MIMO systems, \cite{Zhang141} obtained a closed-form approximation and \cite{Wu171} provided a lower bound of the achievable uplink rate with imperfect CSI, respectively. Also, \cite{Sanguinetti17} considered a similar scenario as \cite{Zhang141} and pursued an asymptotic analysis of the achievable rate. Recently, \cite{Ozdogan18,Ozdogan182} systematically investigated the SE performance  with spatially correlated Ricean fading channels and obtained closed-form expressions of uplink/downlink SE for different channel estimation methods. However, it appears that those expressions are untraceable to gain a very clear insight. Also, the corresponding asymptotic analysis in massive antenna regime is obtained at the cost of  the tougher conditions of the spatial covariance matrix and LOS component. From the above discussion, it becomes apparent that a straightforward and exact theoretical analysis of massive MIMO systems with Ricean fading is missing from the open literature.

In this paper, we firstly introduce a general analytic framework for investigating the achievable uplink SE with a linear maximal-ratio combining (MRC) detector. Then, we derive two exact closed-form expressions of the effective signal-to-interference-plus-noise ratio (SINR) based on the LS and MMSE estimation methods, respectively, which apply for any number of antennas $M$  and any Ricean $K$-factor. These expressions are particularly tractable for admitting fast and effective computation. When the Ricean $K$-factor is equal to zero, i.e., Rayleigh fading, our results are substantially simplified. Also, based on the proposed expressions, we investigate in detail the performance of the effective SINR precisely by making either the $M$ or the Ricean $K$-factor becomes infinite. In both cases, it is shown that the effect of  pilot contamination cannot be removed for both LS and MMSE estimation methods; however, if both the $M$ and Ricean $K$-factor continue to grow unbound, the  pilot contamination issue is eliminated only for the MMSE estimation method. Finally, a set of Monte-Carlo simulations is  conducted to validate the the above mentioned analytical results.

\textit {Notation:} Lower-case  and upper-case  boldface letters denote vectors and matrices, respectively; ${\mathbb{C}}^{M\times N}$ denotes the $M\times N$ complex space;   ${\bf A}^{\text T}$, ${\bf A}^{\dag}$, and ${\bf A}^{-1}$ denote  the transpose, the Hermitian transpose, and the inverse of the matrix ${\bf A}$, respectively; ${\bf I}_{M}$ denotes an $M\times M$ identity matrix and ${\bf 0}_{M\times N}$ denotes an $M\times N$ zero matrix. The expectation operation is $\mathbb{E}\{\cdot\}$.   A complex Gaussian random vector ${\bf x}$ is denoted as ${\bf x}\sim\mathcal {C}\mathcal {N}(\bar{{\bf x}},{\bf {\Sigma}})$, where the mean vector is $\bar{{\bf x}}$ and the covariance matrix is ${\bf {\Sigma}}$, while $\|\cdot\|_{2}$ denotes the 2-norm of a vector. Finally, ${\rm {diag}}\left({\bf a}\right)$ denotes a diagonal matrix where the main diagonal entries are the elements of vector ${\bf a}$.

\section{System Model}
In this paper, we consider a typical uplink cellular communication system with $L$ hexagonal cells. Each cell contains a BS and $N$ single-antenna users. Each BS has a uniform linear array with $M$ ($M\gg N$) antennas. In the following, the time-division duplex (TDD) mode is adopted and we assume that the BSs and users in this system are perfectly synchronized. For each channel use, the $M\times 1$ received signal vector of the BS in cell $j$ is given by
\vspace{-2mm}
\begin{align}
 {\bf y}_{j}=\sqrt{\rho_{u}}\sum\limits_{l=1}^{L}{\bf H}_{jl}{\bf x}_{l}+{\bf n}_{j},\tag{1}
\end{align}
where ${\bf x}_{l}$ denotes the $N\times 1$ vector containing the transmitted signals from all the users in cell $l$, which satisfies $\mathbb{E}\left\{{\bf x}_{l}\right\}={\bf 0}_{N\times 1}$ and $\mathbb{E}\left\{{\bf x}_{l}{\bf x}_{l}^{\dag}\right\}={\bf I}_{N}$, $\rho_u$ is the average transmitted power of each user, and ${\bf n}_{j}\in \mathbb{C}^{M\times 1}\sim\mathcal {C}\mathcal {N}({\bf 0}_{M\times 1},{\bf I}_{M})$ is the additive white Gaussian noise (AWGN) vector.

Also, ${\bf H}_{jl}\in\mathbb{C}^{M \times N}$ represents the channel matrix between the  users in cell $l$ and the BS in cell $j$, whose its $n$th column, ${\bf h}_{jln}\in \mathbb{C}^{M\times 1}$, is the channel vector between the user $n$ in cell $l$ and the BS in cell $j$.
Here, the  block-fading model \cite{Ngo13} is utilized where  the large-scale fading coefficients are kept fixed over lots of coherence time intervals and the small-scale fading coefficients remain fixed within a coherence time interval and change between any two adjacent coherence time intervals. Moreover, the large-scale fading coefficients are assumed to be perfectly known at the BS side due to their  slow-varying
nature.  We herein consider the Ricean fading model in \cite{Zhang141} where both a LOS path and a non LOS (NLOS) component exist in the channels between the users and BS in the same cell, while only a NLOS component exists in the channels between the users and BS in different cells. The above mentioned model is reasonable since a LOS component is more likely to kick in when the users and the BS are in the same cell. Hence, ${\bf h}_{jln}$ can be rewritten as
\begin{align}
\begin{split}
    {\bf h}_{jln}\!=\!\left\{
    \begin{array}{*{20}l}
   \!\!\!\!\!\sqrt{\!\frac{K_{jn}}{K_{jn}+1}}{\bf h}_{jjn, {\rm {LOS}}}+\!\!\sqrt{\!\frac{1}{K_{jn}+1}}{\bf h}_{jjn, {\rm {NLOS}}},  &l=j,\\
   \\
   {\bf h}_{jln, {\rm {NLOS}}}, &l\neq j,\\
     \end{array}\right.
 \end{split}\tag{2}
\end{align}
where $K_{jn}$ denotes the Ricean $K$-factor for the user $n$ in cell $j$, ${\bf h}_{jln, {\rm {NLOS}}}\sim\mathcal {C}\mathcal {N}({\bf 0}_{M\times 1},\beta_{jln}{\bf I}_{M})$ ($\forall l,n$) denotes the NLOS component between the user $n$ in cell $l$ and the BS in cell $j$, while $\beta_{jln}$ is the corresponding large-scale fading coefficient. Also,  ${\bf h}_{jjn, {\rm {LOS}}}\in\mathbb{C}^{M \times 1}$ denotes the LOS part between the user $n$ and the BS in cell $j$, whose $m$th entry $[{\bf h}_{jjn, {\rm {LOS}}}]_{m}$ is given by
\begin{align}
[{\bf h}_{jjn, {\rm {LOS}}}]_{m}=\beta_{jjn}^{\frac{1}{2}}e^{-i(m-1)\frac{2\pi d}{\lambda}\sin(\theta_{jn})},\tag{3}
\end{align}
where $\lambda$ is the wavelength,  $d$ is the antenna spacing, $\theta_{jn}\in [0, 2\pi)$ is the angle of arrival of the $n$th user in cell $j$, and $i$ denotes imaginary unit. Based on the fact that $K_{jn}$ and $\theta_{jn}$ can be obtained through a feedback link \cite{Maddah-Ali12}, we can also assume that the Ricean $K$-factor and LOS path can be perfectly obtained by the BS and users.

In practical communication systems, the BS side does not have perfect CSI and the channel needs to be estimated at the BS. In TDD mode, uplink pilot training is adopted for obtaining the estimated channel information before  data transmission. The worst-case of pilot sequence allocation is adopted here, where each cell's users utilize the same set of pilot sequences \cite{Liu17}, which we denote as ${\bm \Phi}\in\mathbb{C}^{\tau\times N}$. Also, $\tau$ is the length of the pilot sequence, which is larger than or equal to $N$, and ${\bm \Phi}$ satisfies ${\bm \Phi}^{\dag}{\bm \Phi}={\bf I}_{N}$. Then, the $M\times \tau$ received pilot sequence signal matrix at the BS in cell $j$ is given by
\begin{align}
 {\bf Y}_{j,\text {train}}=\sum\limits_{l=1}^{L}{\bf H}_{jl}({\bm \Omega}_{l}+{\bf I}_{N})^{\frac{1}{2}}{\bf P}_{l}^{\frac{1}{2}}{\bm \Phi}^{\dag}+{\bf N}_{j},\tag{4}
\end{align}
where ${\bf P}_l$ is the $N\times N$ diagonal matrix with the power of the pilot sequence sent by user $n$ in cell $l$ is $[{\bf P}_{l}]_{nn}=\rho_{ln}$, ${\bf N}_{j}\in \mathbb{C}^{M\times \tau}$ is the AWGN matrix with  the independent
and identically distributed zero-mean and unit-variance elements, and ${\bm \Omega}_{l}={\rm {diag}}([K_{l1},\ldots, K_{ln},\ldots,K_{lN}]^{\text T})\in \mathbb{C}^{N\times N}$. Particularly,  the reason for the pilot matrix ${\bf P}_{l}^{\frac{1}{2}}{\bm \Phi}^{\dag}$ multiplied by $({\bm \Omega}_{l}+{\bf I}_{N})^{\frac{1}{2}}$, which is sent by all users in cell $j$, is to estimate the NLOS component in a  simplified manner; note that a similar method followed in \cite{Zhang14,Zhang141}. By removing the known LOS part in (4) and utilizing the LS and MMSE estimation methods \cite{Liu17,Kay93},   ${\bf h}_{jjn, {\rm {NLOS}}}$ is estimated as
\begin{align}
{\bf {\hat h}}_{jjn, {\rm {NLOS}}}^{\text {LS}}
&=
{\bf {h}}_{jjn, {\rm {NLOS}}}
\!+\!\!\!\sum\limits_{l\neq j}^{L}\!\!\sqrt{\frac{\rho_{ln}\!(\!K_{ln}\!\!+\!\!1\!)}{\rho_{jn}}}{\bf {h}}_{jln, {\rm {NLOS}}}
\!+\!\frac{\tilde{\bf n}_{jn}}{\sqrt{\rho_{jn}}},\tag{5}\\
{\bf {\hat h}}_{jjn, {\rm {NLOS}}}^{\text {MMSE}}&=\chi_{jn}{\bf {\hat h}}_{jjn, {\rm {NLOS}}}^{\text {LS}},\tag{6}
\end{align}
respectively, where ${\bf {\hat h}}_{jjn, {\rm {NLOS}}}^{\text {LS}}$ and ${\bf {\hat h}}_{jjn, {\rm {NLOS}}}^{\text {MMSE}}$ denote the estimators of ${\bf h}_{jjn, {\rm {NLOS}}}$ based on the LS and MMSE estimation methods, respectively, as well as, $\tilde{\bf n}_{jn}\triangleq{\bf N}_{j}{\bm \phi}_{n}$. Also, ${\bm \phi}_{n}$ is the $n$th column of ${\bm \Phi}$ and
\begin{align}
\chi_{jn}\triangleq\frac{\rho_{jn}\beta_{jjn}}{\rho_{jn}\beta_{jjn}+\sum\limits_{l\neq j}^{L}{\rho_{ln}}(K_{ln}+1)\beta_{jln}+1}.\tag{7}
\end{align}
For convenience,  we denote the estimator of ${\bf {h}}_{jjn, {\rm {NLOS}}}$ as ${\bf {\hat h}}_{jjn, {\rm {NLOS}}}$ and the estimator of ${\bf {h}}_{jjn}$ as
\begin{align}
{\bf {\hat h}}_{jjn}=\sqrt{\frac{K_{jn}}{K_{jn}+1}}{\bf h}_{jjn, {\rm {LOS}}}+\sqrt{\frac{1}{K_{jn}+1}}{\bf {\hat h}}_{jjn, {\rm {NLOS}}},\tag{8}
\end{align}
 respectively, for both the LS and MMSE estimation methods.

After channel estimation, we use the  standard linear detector MRC \cite{Liu17} to detect the received data signal in (1).
Hence, for the BS in cell $j$, (1) is separated into $N$ streams by multiplying it with the MRC detector, that is,
\begin{align}
     {\bf r}_{j}=\hat{\bf H}_{jj}^{\dag}{\bf y}_{j}\in\mathbb{C}^{N \times 1},\tag{9}
\end{align}
where $\hat{\bf H}_{jj}\triangleq[\hat{\bf h}_{jj1},\ldots,\hat{\bf h}_{jjn} ,\ldots,\hat{\bf h}_{jjN}]\in\mathbb{C}^{M \times N}$. Then, for the $n$th user in cell $j$,  we have
\begin{align}
{r}_{jn}&=\sqrt{\rho_{u}}\hat{\bf h}_{jjn}^{\dag}{\bf h}_{jjn}x_{jn}
+\sqrt{\rho_u}\!\!\!\!\sum\limits_{(l,t)\neq(j,n)}\!\!\!\!\hat{\bf h}_{jjn}^{\dag}{\bf h}_{jlt}x_{lt}
+\hat{\bf h}_{jjn}^{\dag}{\bf n}_{j},\tag{10}
\end{align}
where ${r}_{jn}$ is the $n$th entry of ${\bf r}_{j}$.

\section{Spectral Efficiency}
In this section, we obtain the closed-form expressions of the effective SINR for both LS and MMSE estimation methods and, thereafter, aim to analyze the SINR performance with the respect of the number of BS antennas $M$ and the Ricean $K$-factor, respectively.

\subsection{Closed-Form of ${\text {SINR}}_{jn}$}
Since we want to obtain a computable expression of the achievable uplink SE  and investigate it by a simple way, it is convenient to follow the methodology of \cite{Liu17} that assumes that the term $\mathbb{E}\{{\bf \hat{h}}_{jjn}^{\dag}{\bf h}_{jjn}\}$ is known at the BS in cell $j$ perfectly. Hence, (10) can be rewritten as
\begin{align}
     {r}_{jn}&=\!\!\sqrt{\rho_{u}}\mathbb{E}\left\{\hat{\bf h}_{jjn}^{\dag}{\bf h}_{jjn}\right\}x_{jn}
              \!+\!
              \sqrt{\rho_u}\!\!\!\!\!\!\sum\limits_{(l,t)\neq(j,n)}\!\!\!\!\!\!\hat{\bf h}_{jjn}^{\dag}{\bf h}_{jlt}x_{lt}+\sqrt{\rho_{u}}\left\{\hat{\bf h}_{jjn}^{\dag}{\bf h}_{jjn}\!\!-\!\!\mathbb{E}\left\{\hat{\bf h}_{jjn}^{\dag}{\bf h}_{jjn}\!\right\}\right\}\!x_{jn}
              \!+\!\hat{\bf h}_{jjn}^{\dag}{\bf n}_{j}.\tag{11}
\end{align}
Then, by using the definition of the effective SINR in multi-cell massive MIMO systems as in \cite[Eq. (18)]{Liu17}, the achievable uplink SE of the $n$th user in cell $j$, in units of bit/s/Hz, is given by
\begin{align}
    {R}_{jn}=\frac{T-\tau}{T}\log_{2}\left(1+{\text {SINR}}_{jn}\right),\tag{12}
\end{align}
where $T$ denotes the  channel coherence time interval, in terms of the number of symbols, while $\tau$ symbols are utilized for channel estimation, and the ${\text {SINR}}_{jn}$ is defined as
\begin{align}
{\text {SINR}}_{jn}\triangleq\frac{\rho_{u}\left|\mathbb{E}\left\{\hat{\bf h}_{jjn}^{\dag}{\bf h}_{jjn}\right\}\right|^{2}}{\rho_{u} \sum\limits_{l=1}^{L}\sum\limits_{t=1}^{N} \mathbb{E} \left\{\left|\hat{\bf h}_{jjn}^{\dag}{\bf h}_{jlt}\right|^{2}\right\}-\rho_{u}\left|\mathbb{E}\left\{\hat{\bf h}_{jjn}^{\dag}{\bf h}_{jjn}\right\}\right|^{2}+\mathbb{E}\left\{\left\|\hat{\bf h}_{jjn}\right\|_{2}^{2}\right\}}.\tag{13}
\end{align}

The following theorem presents a new general framework for the  closed-form expression of ${\text {SINR}}_{jn}$, which applies for both the LS and MMSE estimation methods. This constitutes a key contribution of this paper.

{\emph {Theorem 1:}} The exact ${\text {SINR}}_{jn}$, for both the LS and MMSE estimation methods, can be analytically evaluated as
\begin{align}
\begin{split}
   {\text {SINR}}_{jn}=\left\{
\begin{array}{*{20}l}
   {\text {SINR}}_{jn}^{\rm {LS}}, &{\text {LS}},\\
   \\
   {\text {SINR}}_{jn}^{\rm {MMSE}}, &{\text {MMSE}},\\
\end{array}\right.
\end{split}\tag{14}
\end{align}
where
\begin{align}
{\text {SINR}}_{jn}^{\rm {LS}}
&\triangleq
    \frac{M\rho_{jn}(K_{jn}+1)\beta_{jjn}^2}
    {M\psi_{jn}
    +\zeta_{jn}\vartheta_{j}
    +\rho_{jn}K_{jn}\beta_{jjn}\varsigma_{jn}},\tag{15}\\
{\text {SINR}}_{jn}^{\rm {MMSE}}
&\triangleq\frac{M{\rho_{jn}}\left(K_{jn}\!+\!\chi_{jn}\right)^{2}\beta_{jjn}^{2}}
    {M\chi_{jn}^{2}\!\left(\!K_{jn}\!+\!1\right)\psi_{jn}
    \!+\!\rho_{jn}\left(K_{jn}\!+\!\chi_{jn}\right)\left(K_{jn}\!+\!1\right)\beta_{jjn}\vartheta_{j}
    \!+\!\rho_{jn}K_{jn}\left(K_{jn}\!+\!1\right)\beta_{jjn}\varsigma_{jn}}.\tag{16}
\end{align}
Also, $\psi_{jn}$, $\zeta_{jn}$, $\vartheta_{j}$, and $\varsigma_{jn}$ are denoted as
\begin{align}
    \psi_{jn}&\triangleq\sum\limits_{l\neq j}^{L}\rho_{ln}\left(K_{ln}+1\right)\beta_{jln}^2,\tag{17}\\
    \zeta_{jn}&\triangleq\sum\limits_{c=1}^{L}\rho_{cn}\left(K_{cn}+1\right)\beta_{jcn}+1,\tag{18}\\
    \vartheta_{j}&\triangleq\sum\limits_{l=1}^{L}\sum\limits_{t=1}^{N}\beta_{jlt}+\frac{1}{\rho_u},\tag{19}\\
    \varsigma_{jn}&\triangleq\sum\limits_{t\neq n}^{N}\frac{K_{jt}}{K_{jt}+1}\frac{\phi_{nt}^2}{M}\beta_{jjt}
    -\sum\limits_{t=1}^{N}\frac{K_{jt}}{K_{jt}+1}\beta_{jjt},\tag{20}
\end{align}
respectively, where
\begin{align}
    \phi_{nt}&\triangleq\frac{\sin\left(\frac{M\pi}{2}\left(\sin(\theta_{jn})-\sin(\theta_{jt})\right)\right)}
    {\sin\left(\frac{\pi}{2}\left(\sin(\theta_{jn})-\sin(\theta_{jt})\right)\right)}.\tag{21}
\end{align}

{\emph {Proof:}} See Appendix A. \hfill\rule{3mm}{3mm}

It is important to note that the expressions in {\emph {Theorem 1}} can be easily evaluated since they involve only the pilot sequence power, uplink data power, Ricean $K$-factor, and large-scale fading coefficients, for all cases of interest. Moreover, from {\emph {Theorem 1}}, we see that the obtained effective SINRs based on the LS and MMSE channel estimation methods with MRC detector are different. Note that when $K_{ln}=0\ (\forall l,n)$, ${\text {SINR}}_{jn}$ reduces to the special case of Rayleigh fading channel. After performing some simplifications,
for both LS and MMSE estimation methods, we have
\begin{align}
&{\text {SINR}}_{{\rm {Rayleigh}},jn}=
\frac{M\rho_{jn}\beta_{jjn}^{2}}{M\sum\limits_{l\neq j}^{L}\rho_{ln}\beta_{jln}^{2}
\!+\!\left(\sum\limits_{c=1}^{L}\rho_{cn}\beta_{jcn}
+1\!\right)\vartheta_{j}}.\tag{22}
\end{align}
Interestingly, (22) is the effective SINR in Rayleigh fading channels given by \cite[Theorem 1]{Liu17}.
Note that {\emph {Theorem 1}} gives a universal formula for the ${\text {SINR}}_{jn}$ when Ricean fading is considered.

\subsection{Analysis of ${\text {SINR}}_{jn}$}
Now, we consider the ${\text {SINR}}_{jn}$ limit when $M$ grows without bound. To the best of our knowledge, this result is also new.

{\emph {Corollary 1:}} If $M\rightarrow\infty$, the exact analytical expression of the ${\text {SINR}}_{jn}$ in (14) approaches to
\begin{align}
\begin{split}
   \lim\limits_{M\rightarrow\infty}{\text {SINR}}_{jn}=\left\{
\begin{array}{*{20}l}
   \overline{{{\text {SINR}}}}_{jn}^{\rm {LS}}, &{\text {LS}},\\
   \\
   \overline{{\text {SINR}}}_{jn}^{\rm {MMSE}}, &{\text {MMSE}},\\
\end{array}\right.
\end{split}\tag{23}
\end{align}
where
\begin{align}
\overline{{\text {SINR}}}_{jn}^{\rm {LS}}&\triangleq
\frac{\rho_{jn}\left(K_{jn}+1\right)\beta_{jjn}^2}
{\psi_{jn}},\tag{24}\\
\overline{{\text {SINR}}}_{jn}^{\rm {MMSE}}&\triangleq
\frac{\rho_{jn}\left(K_{jn}+\chi_{jn}\right)^{2}\beta_{jjn}^2}
{\chi_{jn}^2\left(K_{jn}+1\right)\psi_{jn}}.\tag{25}
\end{align}

{\emph {Proof:}} The proof is completed by calculating the limit of (14) when $M\rightarrow\infty$.
\hfill\rule{3mm}{3mm}

{\emph {Corollary 1}} indicates that if the number of users is kept fixed and the number of receive antennas at the BS side is increased, then, the asymptotic ${\text {SINR}}_{jn}$ is saturated. Intuitively, this is due to the pilot contamination since the other cells' users adopt the same pilot sequence as the user in the target cell $j$. It is also worth noting that, based on (24) and (25), if $K_{ln}=0,\ \forall l,n$, the limit of ${\text {SINR}}_{jn}$ as $M\rightarrow\infty$ for both LS and MMSE estimation methods is given by
\begin{align}
\overline{{\text {SINR}}}_{{\rm {Rayleigh}}, jn}=
\frac{\rho_{jn}\beta_{jjn}^{2}}{\sum\limits_{l\neq j}^{L}\rho_{ln}\beta_{jln}^2}.\tag{26}
\end{align}
Again, it is important to note that (26) is the limit of effective SINR with imperfect CSI in Rayleigh fading channels given by \cite[Corollary 1]{Liu17}. Hence, (26) is a special case of (23) if the power of the LOS part of the Ricean fading channel is equal to zero. To gain more insights, the exact ${\text {SINR}}_{jn}$ admits further simplifications in the large Ricean $K$-factor regime for both LS and MMSE estimation methods.

 {\emph {Corollary 2:}} If for any $l$ and $n$, $K_{ln}=K\rightarrow\infty$, (14) converges to
\begin{align}
\begin{split}
   \lim\limits_{K\rightarrow\infty}{\text {SINR}}_{jn}=\left\{
\begin{array}{*{20}l}
   \widetilde{{\text {SINR}}}_{jn}^{\rm {LS}}, &{\text {LS}},\\
   \\
   \widetilde{{\text {SINR}}}_{jn}^{\rm {MMSE}}, &{\text {MMSE}},\\
\end{array}\right.
\end{split}\tag{27}
\end{align}
where $\widetilde{{\text {SINR}}}_{jn}^{\rm {LS}}$ and $\widetilde{{\text {SINR}}}_{jn}^{\rm {MMSE}}$ are defined as
\begin{align}
\widetilde{{\text {SINR}}}_{jn}^{\rm {LS}}\!&\triangleq\!
\frac{M\!\rho_{jn}\beta_{jjn}^2}
{M\!\!\sum\limits_{l\neq j}^{L}\!\rho_{ln}\beta_{jln}^2
\!\!+\!\!\sum\limits_{c=1}^{L}\!\rho_{cn}\beta_{jcn}\!\vartheta_{j}
+\rho_{jn}\beta_{jjn}\varrho_{jn}},\tag{28}\\
\widetilde{{\text {SINR}}}_{jn}^{\rm {MMSE}}&\triangleq
\frac{M\beta_{jjn}}
{\sum\limits_{l\neq j}^{L}\sum\limits_{t=1}^{N}\beta_{jlt}+\frac{1}{\rho_{u}}
+\sum\limits_{t\neq n}^{N}\frac{\phi_{nt}^{2}}{M}},\tag{29}
\end{align}
respectively. Also, $\varrho_{jn}$ is denoted as
\begin{align}
\varrho_{jn}\triangleq
\sum\limits_{t\neq n}^{N}\frac{\phi_{nt}^{2}}{M}\beta_{jjt}-\sum\limits_{t=1}^{N}\beta_{jjt}.\tag{30}
\end{align}

{\emph {Proof:}}  The proof is completed by calculating the limit of (14) when $K\rightarrow\infty$.
\hfill\rule{3mm}{3mm}

It is interesting to note from {\emph {Corollary 2}} that as $K$ increases, the SINR based on LS and MMSE will approach two constant values, respectively. Note that (29) is unbounded as $M\rightarrow\infty$, whereas (28) is bounded for the same conditions. In other words, in the high Ricean $K$-factor regime with infinite $M$,  the pilot contamination effect will be completely eliminated for MMSE estimation, since
this scheme accounts for the presence of a LOS component in (6). On the other hand,  pilot contamination effect cannot be be removed when using the LS estimation method since the LS estimation method regards the co-channel interference in (5) as just noise. Note that, ${\emph {Corollary 2}}$ reflects that massive MIMO systems have unlimited achievable uplink SE via different estimation methods, the number of BS antennas, and Ricean $K$-factor, though in a slightly different way than in \cite{Bjornson18}.

\subsection{Numerical Results}
In this section, we consider a hexagonal cellular network with a set of $L$ cells and radius (from center to vertex) $\varpi$ meters where users are distributed uniformly in each cell.  Also, leveraging the large-scale fading model of\cite{Ngo141}, the large-scale fading coefficient between the user $n$ in cell $l$ and the BS in cell $j$, $\beta_{jln}$, is given by
\begin{align}
   \beta_{jln}=\frac{v_{jln}}{1+\left(\frac{\eta_{jln}}{\eta_{\rm {min}}}\right)^{\alpha}},\tag{31}
\end{align}
where $v_{jln}$ is a log-normal random variable with standard deviation $\xi$, $\alpha$ is the path loss exponent, $\eta_{jln}$ is the distance between the user $n$ in cell $l$ and the BS in cell $j$, and $\eta_{\rm {min}}$ is the reference distance. In our simulations, we choose $L=7$, $N=10$, $\varpi=500$m, $\xi=8$dB, $\alpha=3.8$, $d=\frac{\lambda}{2}$, and $\eta_{\rm {min}}=200$m, which follow the methodology of \cite{Ngo141, Zhang14}. Here, the observed cell is the center cell and call it cell 1, i.e., $j=1$.
Also, the pilot sequence symbol and the data symbol  are assumed to be modulated based on orthogonal frequency-division multiplexing (OFDM). By considering the long term evolution standard, the channel coherence time interval is equal to 196 OFDM symbols, i.e., $T=196$ \cite{Ngo13,Zhang14}. Since we assume that the noise variance is 1, we consider that each user has the same pilot sequence power denoted by $\rho_p$ which is equal to 30dB, and the uplink data power $\rho_{u}=20$dB. For convenience, we assume all the channels between the BS and the users in same cell have the same Ricean $K$-factor, denoted by $K$, and $\tau=N=10$ OFDM symbols. Finally, all the simulation results are
obtained by averaging 100 realizations of all users' large-scale fading coefficients in all cells over 100 independent small-scale fading channels for each realization of users' large-scale fading coefficients.

In the following, we assess the accuracy of the achievable uplink SE given by (12) for both LS and MMSE estimation methods,  the closed-form expression given in {\emph {Theorem 1}}, and the results in {\emph {Corollary 1}} and {\emph {Corollary 2}}. For comparison, we define the metric called  ``Sum achievable uplink SE" in target cell 1, which is given as
\begin{align}
R_{\rm {sum}}\triangleq\sum\limits_{n=1}^{N}R_{1n}.\tag{32}
\end{align}

\begin{figure*}[!t]
\centering
\subfigure[Moderate-to-High $M$ regime] { \label{fig:(a)}
\includegraphics[width=3.1in]{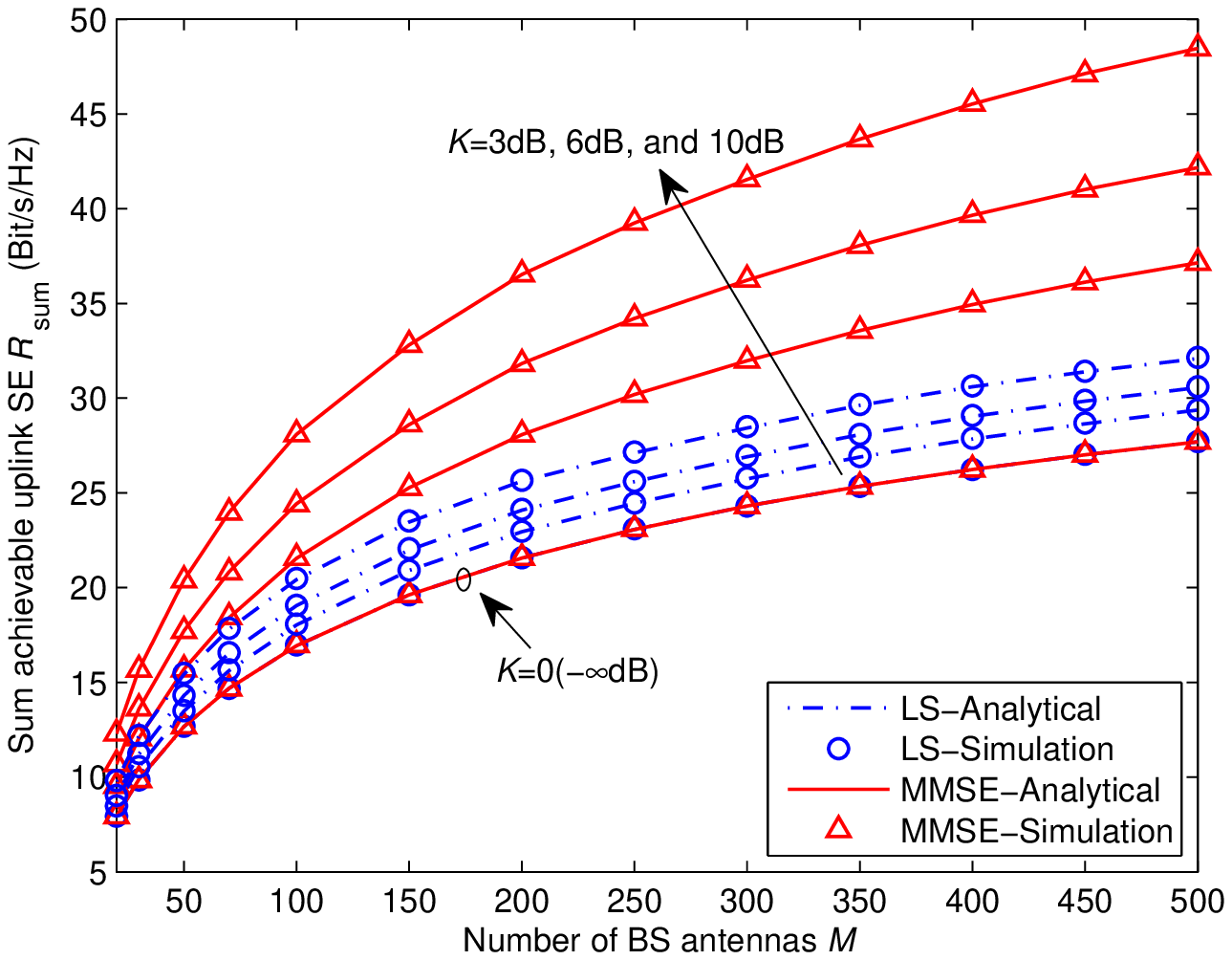}
}
\subfigure[Whole $M$ regime] {\label{fig:(b)}
\includegraphics[width=3.1in]{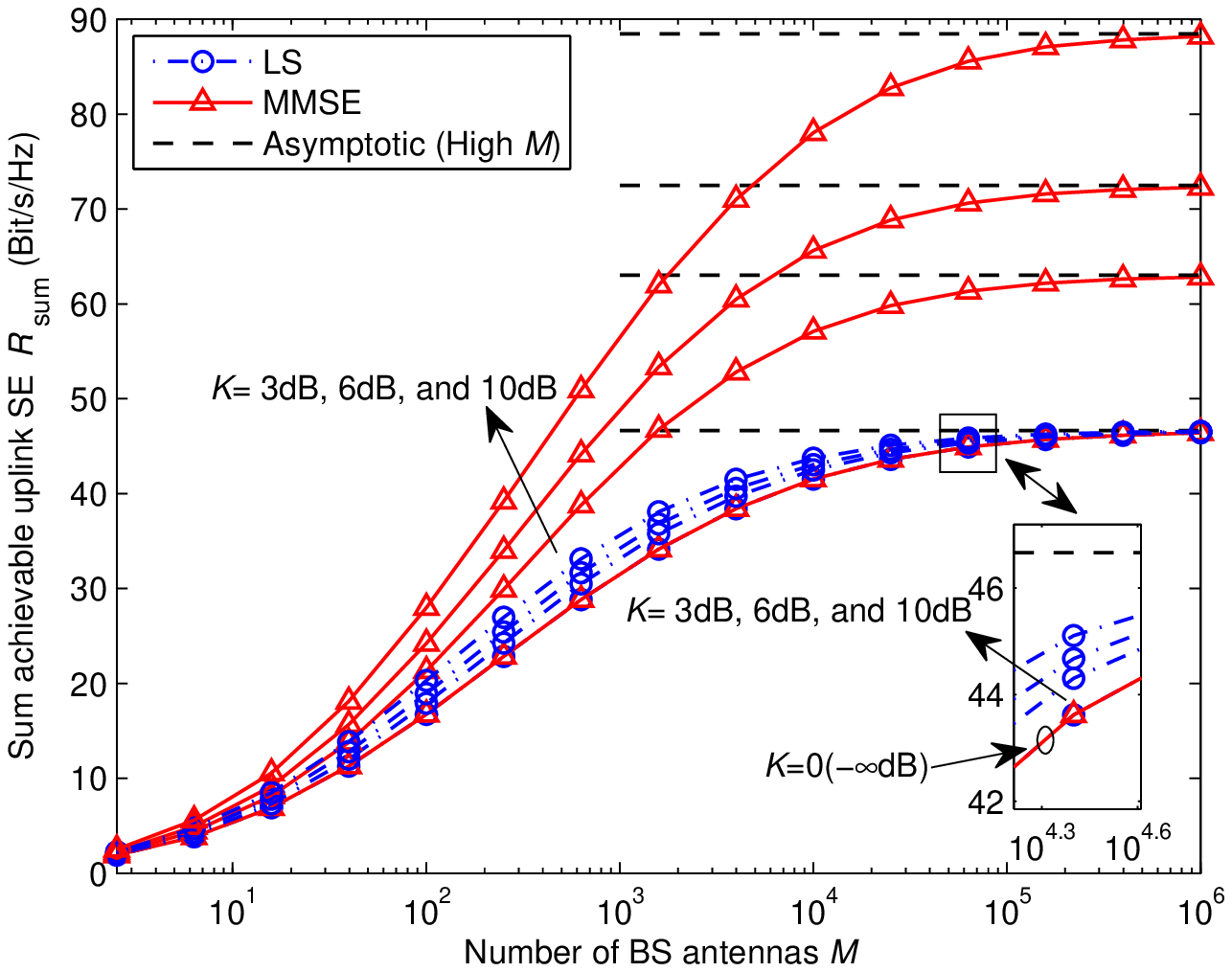}
}
\caption{The sum achievable uplink SE $R_{\rm {sum}}$ as a function of the number of BS antennas $M$ with $\rho_p=30$dB, $\rho_u=20$dB, as well as, $K=0(-\infty{\text{dB}}), 3{\text{dB}}, 6{\text{dB}}, {\text {and}}\ 10{\text{dB}}$, for the LS and MMSE estimation methods, respectively.}
\label{fig}
\end{figure*}

Fig. 1(a) gives the analytical and Monte-Carlo simulated sum achievable uplink SE $R_{\rm {sum}}$ with the LS and MMSE estimation methods, respectively, in moderate-to-high $M$ regime. Results are shown for different Ricean $K$-factor, and pilot sequence power $\rho_p=30$dB with uplink data power $\rho_u=20$dB. We see that in all cases the analytical curves (based on (14)) match precisely with the simulated curves (based on (13)), which proves the validity of {\emph {Theorem 1}}. Moreover, for all cases, when $M$ increases, $R_{\rm {sum}}$ increases. Also, when the power of the LOS path becomes zero, $R_{\rm {sum}}$ for both the LS and MMSE estimation methods are identical for both the analytical and simulated curves, respectively, which not only shows the validity of the results (based on (22)) in previous literature \cite{Liu17}, but also does prove that {\emph {Theorem 1}} can be applied into the Rayleigh fading environment. Moreover, in Ricean fading conditions, the results in this figure show that the MMSE performance is always better than the LS.

Fig. 1(b) investigates the analytical results of sum achievable uplink SE $R_{\rm {sum}}$ in the whole $M$ regime under the same parameter setting in Fig. 1(a).\footnote{Since the match of the analytical results and simulation results has been examined in Fig. 1(a), for convenience, we only need to  examine the analytical results in Fig. 1(b).} When $M\rightarrow\infty$, we see that all $R_{\rm {sum}}$ results tend to different constants, which match the asymptotic expression (based on (23)) for different Ricean $K$-factor, respectively. In other words, it justifies the effectiveness of {\emph {Corollary 1}}. Also, we note that, for the LS case with $K=0(-\infty{\text{dB}}), 3{\text{dB}}, 6{\text{dB}}, {\text {and}}\ 10{\text{dB}}$, as well as, the MMSE case with $K=0(-\infty{\text{dB}})$, the asymptotic results are identical.  This phenomenon is caused by the following two reasons. First, if all users' Ricean $K$-factors are identical, (24)  is uncorrelated with the Ricean $K$-factor. In other words, for the LS case, different Ricean $K$-factor means the same  asymptotic SE. Second, when Ricean $K$-factor is equal to zero, both the LS and MMSE cases have the same asymptotic SE since the current channel becomes the Rayleigh fading channel. Hence,  for both the LS and MMSE estimation methods, the pilot contamination exists such that the $R_{\rm {sum}}$ saturates even when $M\rightarrow\infty$.

\begin{figure}[!t]
\centering
\includegraphics[width=3.1in]{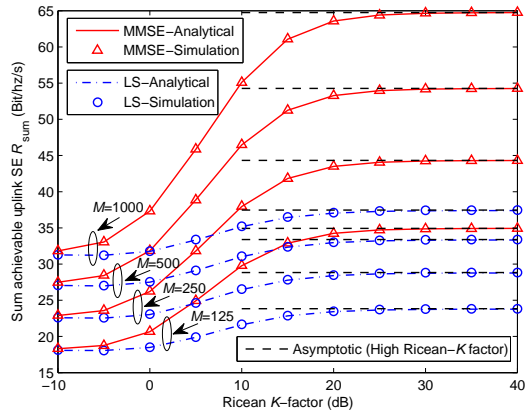}
\caption{The sum achievable uplink SE $R_{\rm {sum}}$ as the Ricean $K$-factor increases with  $\rho_p=30$dB, $\rho_u=20$dB, as well as, $M=125, 250, 500, {\text {and}}\ 1000$, for the LS and MMSE estimation methods, respectively.}
\label{Fig. 1}
\end{figure}

In Fig. 2, we investigate the impact of the Ricean $K$-factor on the sum achievable uplink SE performance for $M=125, 250, 500, {\text {and}}\ 1000$, with the LS and MMSE estimation methods, respectively. In this figure, the pilot sequence power  is  30dB and the uplink data power is  20dB.  It shows that the analytical values and simulation values are almost indistinguishable for both the LS and MMSE estimation methods, regardless  of the number of BS antennas and Ricean $K$-factors. Moreover, across the entire Ricean $K$-factor regime, for both LS and MMSE estimation methods, a larger $M$ means larger $R_{\rm {sum}}$. Given the number of BS antennas $M$, the sum achievable uplink SE performance difference between the LS and and MMSE is distinguishable expect in the low Ricean $K$-factor regime since in this situation the Ricean fading channel tends to become a Rayleigh fading channel.\footnote{In this figure, the asymptotic case when Ricean $K$-factor approaches to $0(-\infty{\text {dB}})$ has not been shown since the Rayleigh fading case has been examined in Fig. 1.} When the Ricean $K$-factor becomes infinite, the sum achievable uplink SE $R_{\rm {sum}}$ approaches to different constant values, which match the asymptotic expressions (based on (27) in {\emph {Corollary 2}}) well, respectively. If also $M\rightarrow\infty$, it can be shown that the pilot contamination is completely removed for the MMSE case.

\section{Conclusion}
In this paper, a detailed statistical characterization of the SE for the muti-cell massive MIMO system with Ricean fading was presented. In order to evaluate the SE performance,
we first proposed two exact closed-form expressions for the effective SINR based on LS and MMSE estimation methods, respectively, which also can be adopted in  Rayleigh fading. Then, we  analyzed the asymptotic properties of the effective SINR when the number of BS antennas $M$ and the Ricean $K$-factor became infinite. It was shown that, when the Ricean $K$-factor became infinite or $M\rightarrow\infty$, the SINR performance for the LS and MMSE estimation methods was saturated, which underlines the pilot contamination phenomenon. However, if both  the Ricean $K$-factor and $M$ grow asymptotically large, the  pilot contamination phenomenon disappeared for the MMSE estimation method, but persists for the LS estimation method.

\appendices

\section{Proof of Theorem 1}
To evaluate the ${\text {SINR}}_{jk}$ in (13), we define six terms
\begin{align}
\mathfrak{A}&\triangleq\left|\mathbb{E}\left\{\hat{\bf h}_{jjn}^{\dag}{\bf h}_{jjn}\right\}\right|^2,\tag{33}\\
\mathfrak{B}&\triangleq\mathbb{E}\left\{\left|\hat{\bf h}_{jjn}^{\dag}{\bf h}_{jjn}\right|^2\right\},\tag{34}\\
\mathfrak{C}&\triangleq\mathbb{E}\left\{\left|\hat{\bf h}_{jjn}^{\dag}{\bf h}_{jjt}\right|^2\right\}, (t\neq n), \tag{35}\\
\mathfrak{D}&\triangleq\mathbb{E}\left\{\left|\hat{\bf h}_{jjn}^{\dag}{\bf h}_{jln}\right|^2\right\}, (l\neq j),\tag{36}\\
\mathfrak{E}&\triangleq\mathbb{E}\left\{\left|\hat{\bf h}_{jjn}^{\dag}{\bf h}_{jlt}\right|^2\right\}, (l\neq j\ \&\ t\neq n),\tag{37}\\
\mathfrak{F}&\triangleq\mathbb{E}\left\{\left\|\hat{\bf h}_{jjn}\right\|_2^2\right\}.\tag{38}
\end{align}

Although $\hat{\bf h}_{jjn}$ has different expressions for the LS and
MMSE estimation methods, the corresponding proofs for ${\text {SINR}}_{jn}$ are similar. Hence,
it is convenient to only study the case of the LS  estimation.
\begin{itemize}
    \item Calculate $\mathfrak{A}$: Substituting (3), (5) and (8) into (33) , after much algebraic manipulation, it can be shown that $\mathfrak{A}$ reduces to
        \begin{align}
        \mathfrak{A}&=M^2\beta_{jjn}^2. \tag{39}
        \end{align}

    \item Calculate $\mathfrak{B}$: Substituting (2), (5), and (8) into (34), after some manipulations, it is easy to obtain
        \vspace{-1mm}
        \begin{align}
        \mathfrak{B}\!&=\underbrace{\!\mathbb{E}\!\left\{\!\left|{\bf h}_{jjn}^{\dag}\!\!\left(\!\!\sum\limits_{l\neq j}^{L}\!\sqrt{\frac{\rho_{ln}\!\left(K_{ln}\!+\!1\right)}{\rho_{jn}}}{\bf h}_{jln,\rm{NLOS}}\!+\!\frac{\tilde{\bf n}_{jn}}{\sqrt{\rho_{jn}}}\!\!\right)\!\right|^2\!\!\right\}}_{B_1}\frac{1}{K_{jn}+1}+\underbrace{\mathbb{E}\left\{\|{\bf h}_{jjn}\|_2^4\right\}}_{B_2}, \tag{40}
        \end{align}
        where the closed-form expression of $B_1$ can be obtained based on ${\bf h}_{jjn}$ is uncorrelated with the rest of terms in $B_1$ and the distribution of ${\bf h}_{jjn}$. Also, $B_2$  can be obtained based on the properties of non-central Wishart matrices
        \cite[Eq. (123)]{Zhang14}. After some algebraic manipulations, we write $\mathfrak{B}$ as follows
        \vspace{-1mm}
        \begin{align}
        \mathfrak{B}&=M^2\beta_{jjn}^2+M\beta_{jjn}^2\frac{K_{jn}}{\left(K_{jn}+1\right)^2}+M\beta_{jjn}\frac{\sum\limits_{l\neq j}^{L}\rho_{ln}\left(K_{ln}+1\right)\beta_{jln}+1}{\rho_{jn}\left(K_{jn}+1\right)}.\tag{41}
        \end{align}
    \item Calculate $\mathfrak{C}$: Since the $\hat{\bf h}_{jjn}$ is uncorrelated with ${\bf h}_{jjt}$ when $t\neq n$,  we substitute (2), (5), and (8) into $\mathfrak{C}$, as well as, utilize  \cite[Eq. (128)]{Zhang14}. Then, clearly
        \begin{align}
        \mathfrak{C}&=\beta_{jjn}\beta_{jjt}\frac{K_{jn}K_{jt}\phi_{nt}^2+M(K_{jn}+K_{jt})+M}{(K_{jn}+1)(K_{jt}+1)}+M\beta_{jjt}\frac{\sum\limits_{l\neq j}^{L}\!\rho_{ln}\!\left(K_{ln}+1\right)\beta_{jln}
                     +1}{\rho_{jn}\left(K_{jn}+1\right)},\tag{42}
        \end{align}
        where $\phi_{nt}$ has been defined as in (21).
    \item Calculate $\mathfrak{D}$: Substituting (2), (5), and (8) into (36), and performing some basic simplifications to obtain
        \begin{align}
         \mathfrak{D}&=\mathbb{E}\left\{\left|D_1\right|^2\right\}
        +\mathbb{E}\left\{\left|D_2\right|^2\right\},\tag{43}
        \end{align}
        where $D_1$  and $D_2$ is denoted as
       \begin{align}
         D_1&\triangleq\!\left(\!{\bf h}_{jjn,\rm{LOS}}\sqrt{\frac{K_{jn}}{K_{jn}\!+\!1}}\!+\!\left(\!{\bf h}_{jjn,\rm {NLOS}}\!+\!\sum\limits_{c\neq j,l}^{L}\sqrt{\frac{\rho_{cn}\!\left(K_{cn}\!\!+\!\!1\right)}{\rho_{jn}}}{\bf h}_{jcn,\rm{NLOS}}\!\!+\!\!\frac{\tilde{\bf n}_{jn}}{\sqrt{\rho_{jn}}}\!\right)\!\!\sqrt{\frac{1}{K_{jn}\!\!+\!\!1}}\right)^{\!\dag}\notag\\
         &\ \ \ \times{\bf h}_{jln,\rm{NLOS}},\tag{44}\\
        D_2&\triangleq\sqrt{\frac{\rho_{ln}\left(K_{ln}+1\right)}{\rho_{jn}\left(K_{jn}+1\right)}}\left\|{\bf h}_{jln,\rm {NLOS}}\right\|_2^2,\tag{45}
        \end{align}
        respectively. Based on the similar way for obtaining $\mathfrak{C}$, $\mathbb{E}\left\{\left|D_1\right|^2\right\}$ reduces to
         \begin{align}
         \mathbb{E}\left\{\left|D_1\right|^2\right\}&=M\beta_{jln}\frac{\sum\limits_{c\neq j,l}^{L}\rho_{cn}\left(K_{cn}+1\right)\beta_{jcn}+1}{\rho_{jn}\left(K_{jn}+1\right)}+M\beta_{jjn}\beta_{jln}.\tag{46}
         \end{align}
         By using the  properties of Wishart matrices \cite[Lemma 2.9]{Tulino04}, we obtain
         \begin{align}
         \mathbb{E}\left\{\left|D_2\right|^2\right\}&=\frac{\rho_{ln}
         \left(K_{ln}+1\right)}{\rho_{jn}\left(K_{jn}+1\right)}
         M\left(M+1\right)\beta_{jln}^2.\tag{47}
         \end{align}
         Therefore, substituting (46) and (47) into (43) and simplifying, we get
         \begin{align}
         \mathfrak{D}&=M^2\beta_{jln}^2\frac{\rho_{ln}\left(K_{ln}+1\right)}
         {\rho_{jn}\left(K_{jn}+1\right)}+M\beta_{jjn}\beta_{jln}+M\beta_{jln}\frac{\sum\limits_{c\neq j}^{L}\rho_{cn}\left(K_{cn}+1\right)\beta_{jcn}+1}
         {\rho_{jn}\left(K_{jn}+1\right)}.\tag{48}
         \end{align}
    \item Calculate $\mathfrak{E}$: Based on the similar way for obtaining $B_1$ and $\mathfrak{C}$, $\mathfrak{E}$ is given by
         \vspace{-0.1in}
        \begin{align}
        \mathfrak{E}&=M\beta_{jjn}\beta_{jlt}+M\beta_{jlt}\frac{\sum\limits_{c\neq j}^{L}\rho_{cn}\left(K_{cn}+1\right)\beta_{jcn}+1}{\rho_{jn}\left(K_{jn}+1\right)}.\tag{49}
        \end{align}
    \item Calculate $\mathfrak{F}$: With the help of (3), (5), and (8), we get
    \vspace{-0.1in}
        \begin{align}
        \mathfrak{F}&=M\beta_{jjn}+M\frac{\sum\limits_{l\neq j}^{L}\rho_{ln}\left(K_{ln}+1\right)\beta_{jln}+1}{\rho_{jn}\left(K_{jn}+1\right)}.\tag{50}
        \end{align}
\end{itemize}
Finally, substituting (39), (41), (42), and (48)-(50)  into (13) and simplifying, the closed-form expression for ${\text {SINR}}_{jn}^{\rm {LS}}$ is obtained. \hfill\rule{3mm}{3mm}

\end{document}